\documentclass{article}

\usepackage[english]{babel}
\usepackage[a4paper,top=2cm,bottom=2cm,left=3cm,right=3cm,marginparwidth=1.75cm]{geometry}
\usepackage{amsmath}
\usepackage{amsfonts}
\usepackage{booktabs}
\usepackage{authblk}
\usepackage{graphicx}
\usepackage{siunitx}
\usepackage{subcaption}
\usepackage{multirow}
\usepackage[colorlinks=true, allcolors=blue]{hyperref}
\mathchardef\mhyphen="2D

\title{
Aluminum solidification and nanopolycrystal deformation via a Graph Neural Network Potential and Million-Atom Simulations
}

\author[1]{Ian Störmer}
\author[1, 2, *]{Julija Zavadlav}
\affil[1]{Professorship of Multiscale Modeling of Fluid Materials, Department of Engineering  Physics and Computation, TUM School of Engineering and Design, Technical University of  Munich, 80333 Munich, Germany}
\affil[2]{Atomistic Modeling Center (AMC), Munich Data Science Institute (MDSI), Technical  University of Munich, 85748 Garching, Germany}
\date{}

\begin{document}
\maketitle
\begin{center}
    *Corresponding Author(s), Email(s): julija.zavadlav@tum.de;
\end{center}

\begin{abstract}
Solidification governs the microstructure and, therefore, the mechanical response of metal components, yet the atomistic details of nucleation and defect formation are often difficult to determine experimentally. Molecular dynamics can bridge this gap, but only if the interatomic model is both accurate and computationally efficient. Here, we develop a Machine Learning Potential (MLP) for aluminum and demonstrate its near ab initio fidelity when trained with the sequential-refinement workflow that fine-tunes the model on low-energy structures. The favorable scaling of the model enables nanosecond simulations involving millions of atoms, thereby overcoming finite-size effects in simulations of polycrystalline solidification and subsequent mechanical testing. Comparison with classical potentials and recent MLP models, including a general-purpose model, shows that inaccuracies in stacking-fault energetics and diffusion can lead to qualitatively incorrect solidified grain structures and post-solidification mechanical behavior. Since our framework is based on an equivariant graph neural network, it allows for straightforward extensions to multi-component systems, providing valuable guidance for the future design and fine-tuning of both specialized and universal MLPs in computational mechanics simulations.
\end{abstract}
\textit{Keywords:}
machine learning interatomic potential, equivariant graph neural network, molecular dynamics, solidification, polycrystals, aluminum

\section{Introduction}
Understanding and controlling the solidification of metals and alloys is crucial to advancing material fabrication through emerging technologies such as additive manufacturing~\cite{debroy2018additive}.
Solidification nucleation occurs at the atomistic scale and cascades to larger scales, influencing the solidification growth and grain microstructure~\cite{curtin2003atomistic}.
These microstructural features, in turn, play a significant role in determining the mechanical properties of the final product, such as ductility, elasticity, and tensile strength~\cite{hall1951deformation, kumar2003mechanical}.
Traditional experimental techniques often require extensive resources and may lack the necessary spatiotemporal resolution~\cite{iqbal2005real, li2006real, zeng2017real}.
As a result, molecular dynamics (MD) simulations have become a valuable complementary tool.
They provide a detailed insight into the onset and evolution of solidification at the per-atom scale, functioning as a computational microscope.

Nevertheless, the predictive capability of MD simulation critically depends on the choice of interatomic potential model.
Classical force fields based on the embedded-atom method (EAM)~\cite{daw1984embedded} and the modified embedded-atom method (MEAM)~\cite{baskes1992modified} have been widely adopted for solidification studies, as they facilitate large-scale simulations that overcome finite-size effects~\cite{mahata2019effects, mahata2019evolution, mahata2018understanding, chang2024formation, mahata2019size}.
Despite their efficiency and robustness, extending these models beyond elemental and binary systems is challenging and labor-intensive.
As a result, reliable classical models are lacking for many industrially relevant multi-component alloys. Moreover, the fixed formulation of the potential energy limits the transferability; models often struggle to predict well properties outside their targeted scope~\cite{becker2013considerations, hale2018evaluating, castillo2022transferability}.
For instance, accurate modeling of both liquid and solid states remains a significant challenge~\cite{castillo2022transferability, mendelev2003development}.

Emerging machine learning potentials (MLPs) offer a promising solution to interatomic modeling~\cite{botu2015learning}.
By integrating the speed of classical force fields with the accuracy of density functional theory (DFT), MLPs enable large-scale simulations with near ab-initio accuracy.
Several MLPs have previously been developed for aluminum due to its widespread use and favorable properties, such as a high strength-to-weight ratio~\cite{stojanovic2018application}.
Notably, studies by Smith et al.~\cite{smith2021automated}, Akhmerov et al.~\cite{akhmerov2024neural}, and Jakse et al.~\cite{jakse2023machine} have shown that MLPs can substantially outperform traditional empirical potentials.
However, these approaches utilized the Behler-Parinello neural network architecture, whose computational complexity scales quadratically with the number of chemical elements, significantly limiting their applicability to multi-element systems such as high-entropy alloys~\cite{behler2007generalized, artrith2017efficient, kyvala2023optimizing}.
Additionally, the explicit calculation of three-body features further raises computational and memory demands.
Thus, while some Behler-Parinello-type and similar MLPs have been used to investigate the solidification of metals or alloys, these studies typically involved small systems~\cite{sun2023molecular} or deep undercooling conditions~\cite{jakse2023machine}, where solidification can be observed already on the picosecond timescale.

To mitigate the unfavorable scaling with species count and the reliance on hand‑crafted descriptors, state‑of‑the‑art models adopt Graph Neural Networks Potentials (GNNs)~\cite{batzner20223, batatia2022mace, musaelian2023learning}, in which atoms are represented as nodes and interatomic interactions are computed by passing messages along edges to neighbors within a cutoff.
Because element identity is encoded via compact learned embeddings and a single, shared set of message‑passing and update functions is applied across all nodes and edges, the computational cost scales primarily with the number of neighbors within a specified cutoff rather than with the number of species.
This weight sharing enables the construction of foundational MLPs, general‑purpose models pretrained on broad materials corpora~\cite{wood2025family, batatia2025foundation}.
For instance, the best‑performing entries on the Matbench Discovery leaderboard~\cite{riebesell2025framework} for diverse solid‑state properties are trained on large‑scale datasets such as Open Materials 2024 (OMat24, $\approx$110 million data points)~\cite{barroso2024open}.
While many GNNs stack multiple message‑passing iterations to expand the receptive field beyond the cutoff, such architectures are difficult to parallelize efficiently across multiple GPUs in domain decomposition implementations.
Consequently, strictly local designs such as Allegro~\cite{musaelian2023learning} are better suited to solidification simulations where larger systems necessitate multi‑GPU execution due to extensive memory needs~\cite{kozinsky2023scaling, fuchs2025chemtrain-deploy}.
Despite their zero‑shot aspiration, foundational models typically still require fine‑tuning to match the accuracy of specialized models~\cite{mannan2025evaluating}, and their inference cost can remain prohibitive even with pruning~\cite{kong2025scalable}.
Accordingly, there remains a need for MLPs that employ modern architectures and are well-suited for large-scale metal solidification simulations.

In this work, we present an MLP for pure aluminum based on the Allegro GNN architecture (GNNP-Al) and apply it to investigate both the solidification process and the mechanical deformation of solidified nanopolycrystals.
By focusing on an elemental metal rather than alloy solidification, we facilitate direct comparison with multiple classical force fields.
Our results reveal that GNNP-Al consistently surpasses all tested classical potentials at elevated temperatures and demonstrates that inadequate modeling of liquid-state properties can, in extreme cases, result in amorphous solidification.
Moreover, GNNP-Al outperforms the small universal model for atoms (UMA-S)~\cite{wood2025family} in predicting energy-volume curves and stacking fault energy profile, which are properties central to solidification pathways and plasticity.
In parallel, GNNP‑Al delivers substantially higher computational efficiency, in both simulation speed and memory usage, than prior specialized and foundational MLPs, enabling temperature-quenching and deformation simulations on million-atom systems at nanosecond timescales.
Notably, the predicted properties of solidified nanopolycrystals qualitatively match experimental observations in predicting five-fold twins, underscoring the physical fidelity of GNNP‑Al and the potential to extend our approach to industry-relevant alloy systems.

\section{Methods}

\subsection{Data}

We reuse the ANI-Al dataset~\cite{smith2021automated}, which was generated via the active learning loop of the ANI-Al MLP training.
It consists of 6352 DFT calculations containing between 55 and 249 aluminum atoms, covering a wide range of solid and liquid states.
The calculations were performed using a Perdew-Burke-Ernzerhof (PBE) functional.
Further information regarding the dataset and its generation is provided in the original work by Smith et al.~\cite{smith2021automated}.
We identified seven samples as outliers while predicting the reference data using both the UMA~\cite{wood2025family} and ANI-Al~\cite{smith2021automated} potentials.
For each of these, either one or both models' predictions differ by more than three standard deviations compared to the rest of the predictions.
Hence, we removed them from the dataset.

Additionally, we generated a training dataset with low-energy samples. For the three crystal structures, face-centered cubic (FCC), hexagonal close-packed (HCP), and body-centered cubic (BCC), we constructed 30 samples close to ideal crystals with varying lattice parameters.
We refer to the new dataset of 90 samples as the low-energy (LE) dataset.
Since the ANI-Al potential matches DFT in this range (Figure \ref{fig:crystal}a), we use it to produce the energy and force labels.
Combined with the original ANI-Al dataset, we refer to the full dataset as ANI-Al+.

\subsection{GNNP Training via Sequential Refinement}

We use the \texttt{chemtrain}~\cite{fuchs2025chemtrain} framework to train our models, as it readily allows us to apply the trained GNNPs to large-scale parallelized simulation in LAMMPS~\cite{thompson2022lammps} via \texttt{chemtrain-deploy}~\cite{fuchs2025chemtrain-deploy}. 
As we discuss in Section \ref{sec:solid}, models trained solely on the ANI-Al dataset underperformed on low-energy states.
This result is not surprising given that the active learning loop primarily explored the high-temperature regime where ideal crystals are rare.
Note that even when samples were differently weighted, we were unable to simultaneously achieve sufficient accuracy on both high- and low-energy samples (Supplementary Note 1). 

\begin{figure}[!htbp]
    \centering
    \includegraphics[width=0.8\linewidth]{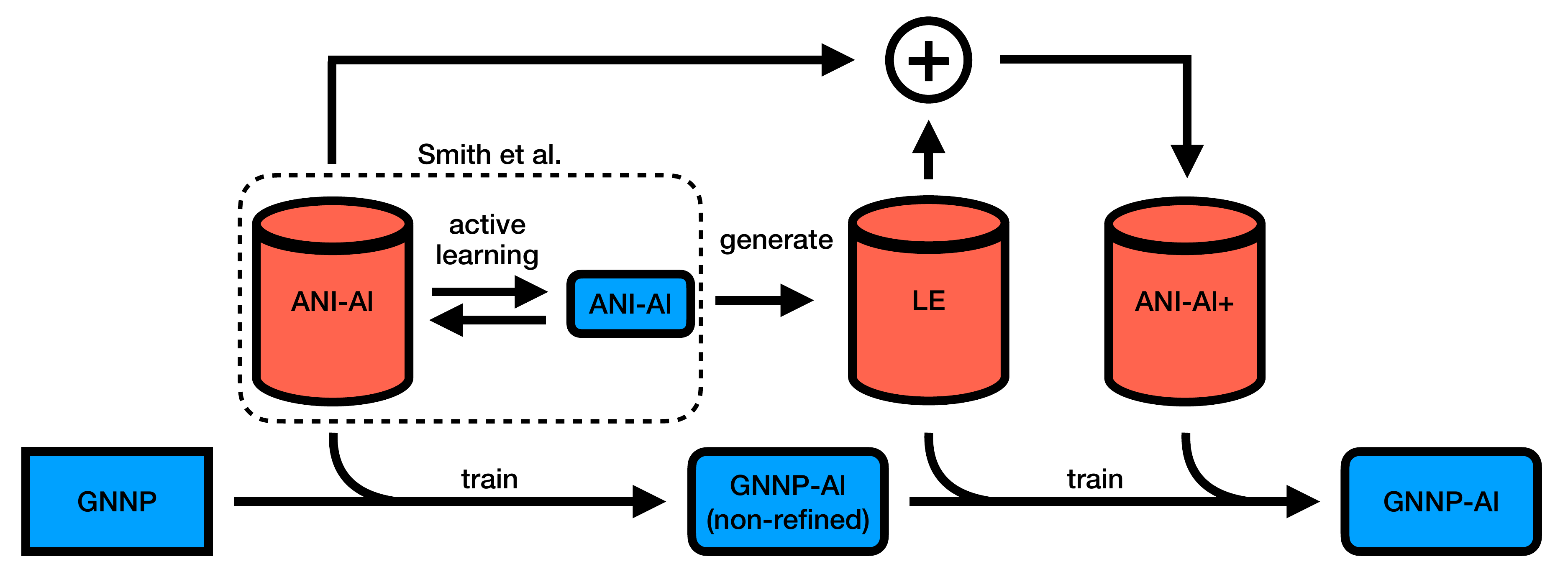}
    \caption{Sequential refinement workflow in three stages. First, the GNNP is pre-trained on the ANI-Al dataset, generated via an active learning loop during training of the corresponding ANI-Al potential. We refer to the resulting model as GNNP-Al (non-refined). Using the ANI-Al model, we generate additional low-energy (LE) samples to refine the pre-trained model in the second stage. In the last step, the refined model is trained on a weighted combination of both datasets, called ANI-Al+, to yield the final GNNP-Al model.
    The components in the dashed box are part of the work by Smith et al.~\cite{smith2021automated}}
    \label{fig:workflow_training}
\end{figure}

To overcome this issue, we devised a new training procedure using sequential refinement, as illustrated in Figure \ref{fig:workflow_training}. 
First, the model is pre-trained on the ANI-Al dataset alone, yielding the potential we refer to as GNNP-Al (non-refined).
In the next step, we refine the model by training it solely on the LE dataset to improve accuracy in low-energy states, such as perfect crystal structures.
Finally, we train on ANI-Al+ with a weighted combination of the two datasets, yielding the final GNNP-Al model. 

We randomly split the ANI-Al dataset~\cite{smith2021automated} into training, validation, and test sets consisting of 70\%, 10\%, and 20\%, respectively.
The resulting subsets are used throughout this work and are not identical to the splits used in Smith et al.~\cite{smith2021automated}.
We only use the LE dataset in training.
The loss function is a weighted sum of the energy and the force mean squared error (MSE), i.e.,
\begin{equation} \label{eq:loss}
        L^{(d)} = \beta_U \sum_{i=1}^{N_d} \left(U_i^{(d)} - \hat{U}_i^{(d)}\right)^2 + \beta_F \sum_{i=1}^{N_d} \frac{1}{3N_i} \sum_{j=1}^{N_i} \left\|\mathbf{F}_{i,j}^{(d)} - \hat{\mathbf{F}}_{i,j}^{(d)}\right\|^2, \quad d \in \{\mathrm{ANI \mhyphen Al}, \mathrm{LE}\},
\end{equation}
where $U_i^{(d)}$ is the reference energy of sample $i$, and $\mathbf{F}_{i,j}^{(d)}$ is the respective force vector acting on atom $j$.
The hat symbol denotes the corresponding model predictions, $N_d$ is the number of samples, and $N_i$ is the number of atoms in sample $i$.
The coefficients $\beta_U$ and $\beta_F$ are the weights of the energy and force MSE, respectively.
During training, contributions from different samples are weighted, depending on the training stage, using
\begin{equation} \label{eq:tot_loss}
    L_{\mathrm{tot}} = w_{\mathrm{ANI \mhyphen Al}} L^{(\mathrm{ANI \mhyphen Al})} + w_{\mathrm{LE}} L^{(\mathrm{LE})}.
\end{equation}
For the first two stages, we set $w_{\mathrm{ANI \mhyphen Al}}=1$ and $w_{\mathrm{LE}}=0$, as well as $w_{\mathrm{ANI \mhyphen Al}}=0$ and $w_{\mathrm{LE}}=1$, respectively, to consider only the corresponding dataset.
In the final training stage, using the combined ANI-Al+ dataset, we chose $w_{\mathrm{ANI \mhyphen Al}}=1$ and $w_{\mathrm{LE}}=10^4$ to compensate for the lower number of samples and ensure the accuracy on this dataset is preserved.

The sequential refinement workflow consistently yielded better results for low-energy states while retaining high accuracy across the dataset, compared to training directly on ANI-Al+ (Supplementary Note 1), most likely because rare low-energy samples are underrepresented even when higher values for $w_{\mathrm{LE}}$ are used. 
We also observed that more epochs are required for the second stage of training.
The training and model parameters are listed in Supplementary Tables 1 and 2, respectively. 

\subsection{Reference Models}
\label{sec:reference_models}

We trained three models using the same sequential refinement method, each with increasing expressivity.
Based on the number of parameters, we label them GNNP-Al-S, M, and L.
The differences between the models are described in Supplementary Note 1 and listed in Supplementary Table 1.
If not stated otherwise, we will use the large GNNP-Al-L as the default model and refer to it as GNNP-Al.
To illustrate the effect of our sequential refinement method, we also include a non-refined GNNP-Al, which is the same GNNP-Al-L pre-trained on ANI-Al, alone.

We reference our GNNP-Al against several previously published interatomic models throughout this study.
This comparison helps identify the strengths of GNNP-Al relative to other MLPs and classical force fields and highlights regimes where some previous models lack the necessary precision. 
For MLPs, we compare to the bespoke ANI-Al model by Smith et al.~\cite{smith2021automated} and the foundational UMA model (UMA-S) by Wood et al.~\cite{wood2025family}.
UMA was trained on several large-scale datasets obtained with different DFT functionals.
We choose the OMat output task or chemical domain such that the model emulates the PBE functional used to generate the OMat24 dataset~\cite{barroso2024open}.
Note that Smith et al.~\cite{smith2021automated} used the same functional to obtain the ANI-Al dataset.

For comparison with classical force fields, we repeat all predictions and simulations using the MEAM potential by Lee et al.~\cite{lee2003semiempirical} as well as three EAM potentials published by Winey et al.~\cite{winey2009thermodynamic}, Mishin et al.~\cite{mishin1999interatomic}, and Mendelev et al.~\cite{mendelev2008analysis}.
These models have been developed for different goals and are therefore fitted to different target properties.
Nevertheless, all models used solid-state properties, such as elasticity, lattice parameters, stacking fault energy, vacancy formation energy, and energy differences between ideal crystal structures, as targets, with reference values obtained either from DFT calculations or experiments.
The EAM model by Mendelev et al.~\cite{mendelev2008analysis} additionally included properties, such as melting temperature, diffusivity, liquid density, and volume and enthalpy changes during melting.
On the other hand, the MEAM model by Lee et al.~\cite{lee2003semiempirical} targeted the thermal expansion coefficient during fitting.
Due to their different approaches, the models exhibit significant differences in performance across different temperature regimes, which, in turn, significantly affect the resulting solidification and deformation simulations (Figure \ref{fig:deformation_ovito} and \ref{fig:deformation_plot}).

\section{Results}

\subsection{Energies and Forces}
\label{sec:energies}
First, we assess how well each model predicts the test set energies and forces on the ANI-Al dataset.
All MLPs achieve mean absolute errors (MAEs) and root mean square errors (RMSEs) for energies below the chemical accuracy threshold of 1~kcal/mol or 43~meV/atom, while the errors of the classical potentials are 1-3 orders of magnitude larger (Table \ref{tab:errors}).
\begin{table}[!htbp]
    \centering
    \begin{tabular}{l r r r r}
        \toprule
        \multirow{2}{*}{Model} & \multicolumn{2}{c}{$U$ [meV/atom]} & \multicolumn{2}{c}{$F$ [meV/Å]} \\
        \cmidrule(lr){2-3}\cmidrule(lr){4-5}
        & MAE & RMSE & MAE & RMSE \\
        \midrule
        GNNP-Al & 6.319 & 7.535 & 70.389 & 94.079 \\
        GNNP-Al (non-refined) & 2.197 & 3.401 & 34.987 & 49.258 \\
        \addlinespace[0.5ex]
        ANI-Al (Smith et al.~\cite{smith2021automated}) & 1.623 & 2.157 & 40.176 & 58.896 \\
        UMA (Wood et al.~\cite{wood2025family}) & 6.078 & 8.108 & 30.258 & 43.605 \\
        \addlinespace[0.5ex]
        MEAM (Lee et al.~\cite{lee2003semiempirical}) & 53.947 & 87.555 & 244.060 & 367.315 \\
        EAM (Winey et al.~\cite{winey2009thermodynamic}) & 130.162 & 252.587 & 513.353 & 1143.814 \\
        EAM (Mishin et al.~\cite{mishin1999interatomic}) & 162.585 & 306.196 & 773.798 & 1885.306 \\
        EAM (Mendelev et al.~\cite{mendelev2008analysis}) & 1889.279 & 14667.087 & 18449.536 & 235159.942 \\
        \bottomrule
    \end{tabular}
    \caption{Mean absolute errors (MAEs) and root mean square errors (RMSEs) for the energy $U$ and force $F$ predictions of different potentials on the ANI-Al test set.}
    \label{tab:errors}
\end{table}
ANI-Al is the most accurate for energy predictions, while UMA performs best for forces.
The energy vs. force accuracy can be, to some extent, prioritized by adjusting the weights $\beta_U$ and $\beta_F$ in Equation \ref{eq:loss}.
Our non-refined GNNP-Al model achieves energy errors comparable to ANI-Al and force errors similar to UMA, demonstrating state-of-the-art capabilities.
Refining the model on additional data, while necessary, slightly reduces accuracy.
It is important to note that ANI-Al has an advantage in this comparison, as the dataset was generated within the active learning loop of this model.
Furthermore, Fu et al.~\cite{fu2022forces} established that energy and force errors alone are insufficient to assess model quality, which we confirm for both MLPs and classical potentials in the following sections.

\begin{figure}[!htbp]
    \centering
    \includegraphics[width=.8\linewidth]{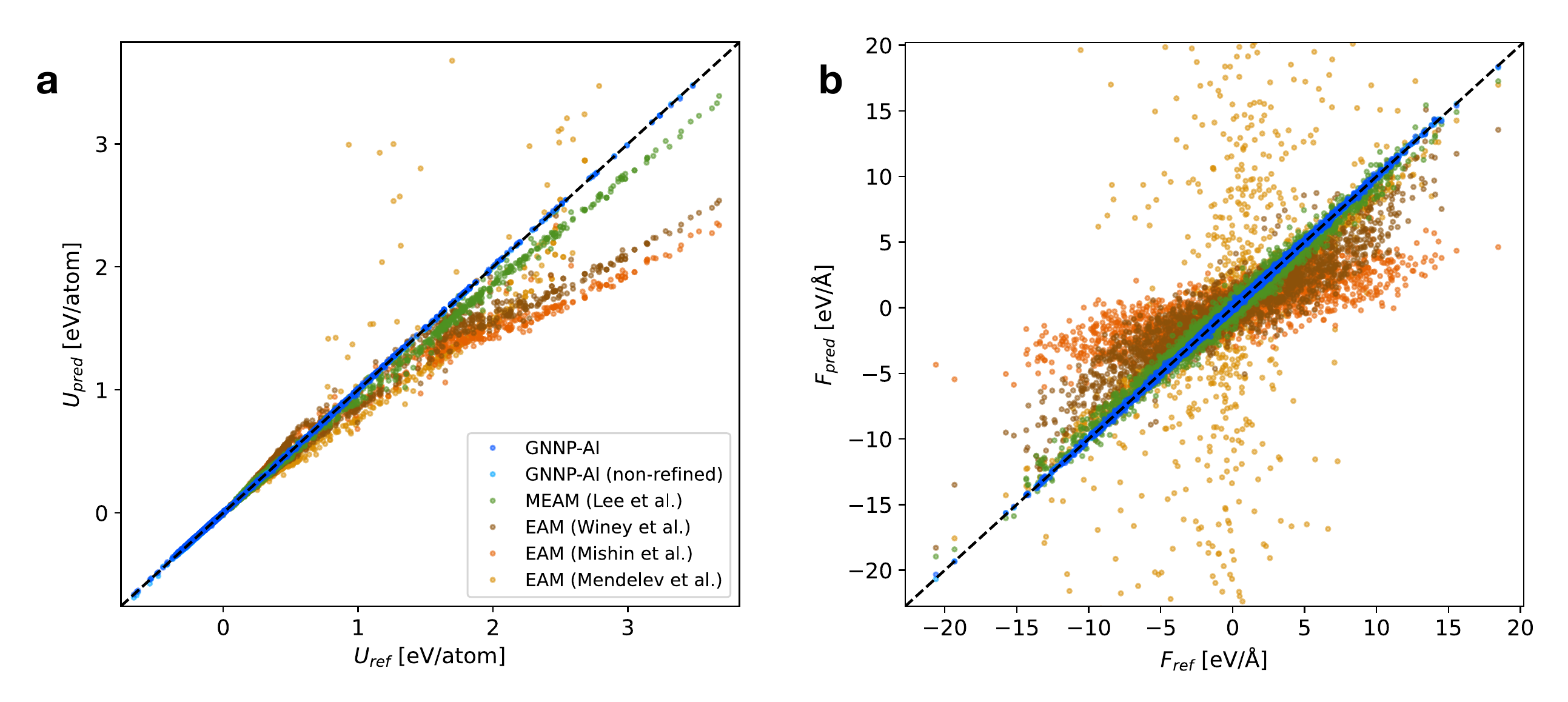}
    \caption{Energies $U$ (a) and force components $F$ (b) predicted by the tested potentials compared to the reference label. The data points correspond to a random subset of the test set of the ANI-Al dataset~\cite {smith2021automated}. We excluded predictions from the ANI-Al~\cite{smith2021automated} and UMA~\cite{wood2025family} models as they are indistinguishable from GNNP-Al. Energies are shifted so that the lowest-energy state is set to zero, enabling comparability across models. Some outliers predicted by the Mendelev et al.~\cite{mendelev2008analysis} model fall outside of the plot range.
    }
    \label{fig:parity}
\end{figure}

The large error metrics for classical potentials stem from the strong influence of high-energy samples. Namely, while all potentials are highly accurate for low-energy samples, corresponding to ordered solid structures, the classical potentials deteriorate at higher energies and force magnitudes, mostly underestimating both (Figure \ref{fig:parity}).
Simpler EAM potentials exhibit this behavior to a larger extent than the more complex MEAM model.
The EAM potential by Mendelev et al.~\cite{mendelev2008analysis} is the only one that shows significant outliers with strongly overestimated energies and forces.
Comparing the drastically larger RMSEs to the MAEs in Table~\ref{tab:errors} underscores the presence of outliers.
Overall, the results indicate that classical potentials model the liquid state poorly, which we investigate further in Section \ref{sec:liquid}.

\subsection{Computational Performance Scaling with System Size}
\label{sec:scaling}

\begin{figure}[!htbp]
    \centering
    \includegraphics[width=.8\linewidth]{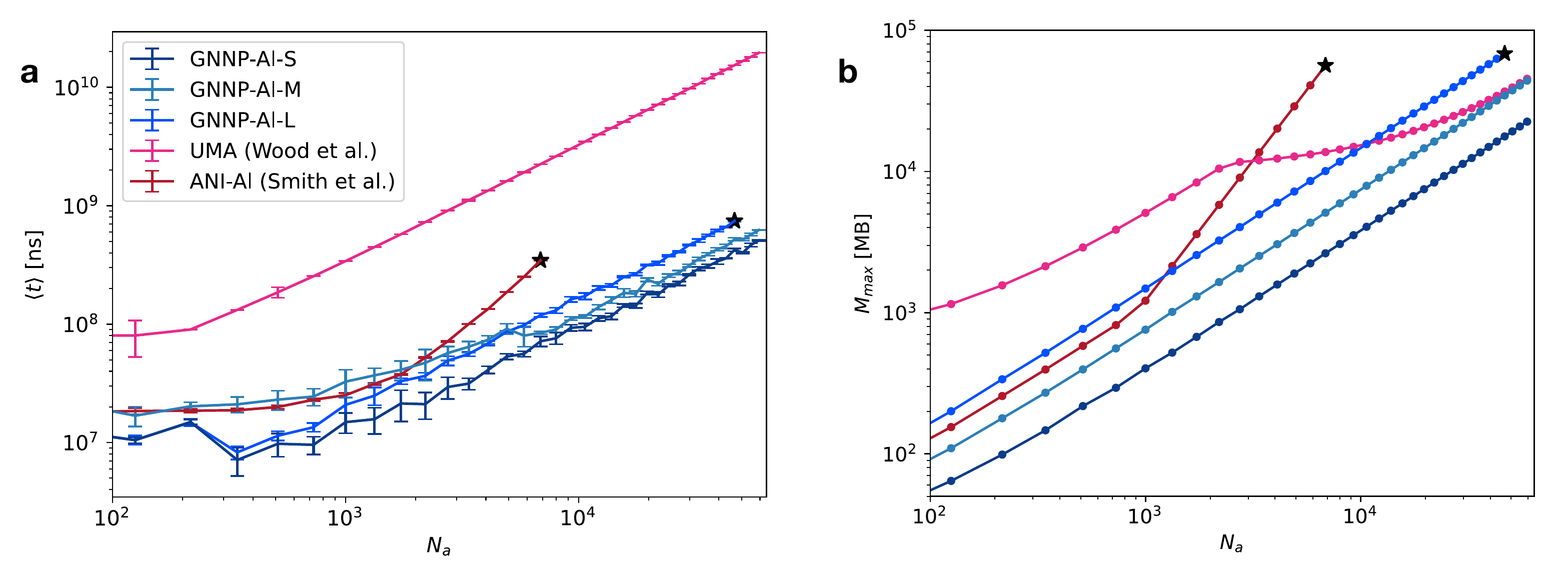}
    \caption{Average inference time $\langle t \rangle$ across 100 runs (a) and peak memory usage $M_{max}$ (b) of the tested MLPs for increasing system sizes, with the number of atoms $N_a$ reaching up to 64000. Error bars represent one standard deviation. The star indicates the largest system size before an out-of-memory error occurred. We performed all tests on a single NVIDIA A100 80GB GPU.}
    \label{fig:scaling}
\end{figure}

To gauge the applicability of different MLPs for solidification simulations, we conduct a scaling test.
Therein, we record the average time and memory required to predict the energy and forces for a single configuration of increasing size (Figure \ref{fig:scaling}).
We perform all tests on the same hardware and using the Atomic Simulation Environment (ASE) software~\cite{hjorth2017atomic}.
For ANI-Al and UMA, we employ the published ASE calculators, whereas for our trained GNNP-Al models, we use equivalent architectures with random weights to avoid retraining in a different environment (Supplementary Note 2).

UMA is an order of magnitude slower than the other MLPs across all system sizes.
At the same time, ANI-Al exhibits quadratic scaling with a rising number of atoms (Figure \ref{fig:scaling}a).
The peak memory usage exhibits similar behavior with an initial offset for UMA and a steeper gradient for ANI-Al (Figure \ref{fig:scaling}b).
Importantly, the ANI-Al model already runs out of memory for a few thousand atoms.
The evaluation of angular symmetry functions requires summations over atomic triplets~\cite{behler2007generalized, kyvala2023optimizing}.
Additionally, the large radial cutoff of $r_c=7$ Å amplifies this problem, leading to $(7/5)^{3} = 2.744$ as many neighbors per atom compared to the $r_c=5$ Å used in many other potentials.
Lastly, ANI-Al consists of an ensemble of 8 neural networks, which are architecturally identical but differently initialized, further increasing memory demands~\cite{smith2021automated}.
UMA, in contrast, employs a smart memory usage strategy, leading to a flattened slope after an initial increase.
To emphasize the severity of the observed scaling, we visualized the same results with linear instead of logarithmic axes in Supplementary Figure 1. 

In contrast, our trained GNNP-Al models scale linearly with system size and remain comparably low in computational cost for all tested systems.
We also observe good memory performance, with solely the large model running out of memory for the largest tested system.
Thus, of the tested models, only our GNNP-Al models are practically feasible for metal solidification simulations and other property predictions, requiring large system sizes and/or long simulations.
Hence, in the following sections, properties such as solidification and nanopolycrystal deformation are compared only with classical potential's predictions.  

\subsection{Solid Properties} 
\label{sec:solid}
\begin{figure}[!htbp]
    \centering
    \includegraphics[width=.8\linewidth]{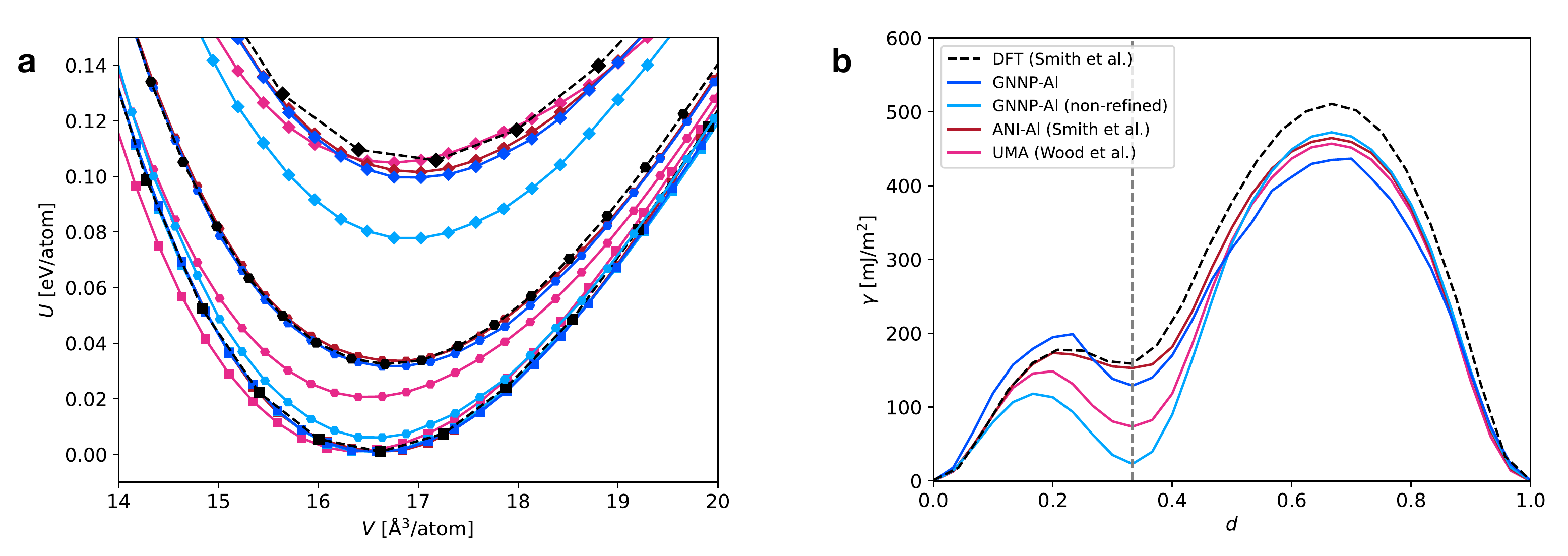}
    \caption{(a) Energies $U$ of the three competing lowest energy crystal structures, face-centered cubic (FCC), hexagonal close-packed (HCP), and body-centered cubic (BCC), with varying per-atom volume $V$. Squares, hexagons, and diamonds denote FCC, HCP, and BCC, respectively. (b) Stacking fault energy $\gamma$ plotted against the normalized displacement $d$ along the fault vector in $\left< 112 \right>$ direction. The dashed gray line at 1/3 corresponds to a stacking fault where FCC locally resembles HCP. The reference density functional theory (DFT) data in both sub-figures were taken from Smith et al.~\cite{smith2021automated}. Predictions by classical potentials are excluded for readability.}
    \label{fig:crystal}
\end{figure}

Next, we analyze the predicted energies for different packing orders and varying lattice parameters (Figure \ref{fig:crystal}a). 
Both GNNP-Al and ANI-Al show good agreement with the DFT reference (see Supplementary Figure 3 for smaller GNNP-Al models).
The relative energy difference between competing crystal structures is among the most important properties for predicting metal solidification.
If the relative energy of a particular crystal lattice is under- or overestimated, it will occur more or less frequently in the solidified polycrystalline material.
To highlight this effect, we include the results from the GNNP-Al (non-refined) model.
GNNP-Al underestimates HCP structure energies and predicts lower energies for BCC packing (Figure \ref{fig:crystal}a).
These errors propagate in solidification simulations, leading to an overrepresentation of the HCP lattice and an increase in the number of grains (Figure \ref{fig:solidification_ovito} and \ref{fig:solidification_plot}).
The non-refined GNNP-Al model also highlights the effectiveness of our sequential refinement methodology in correcting the erroneous energy-difference predictions.
Interestingly, the UMA model shows a similar underestimation of HCP energies.
Thus, while UMA is computationally too expensive for million-atom solidification simulations, it would likely also yield polycrystalline structures with inaccurate HCP-to-FCC lattice fractions. 
Classical models agree well with the DFT reference data for FCC and HCP energies at the minima, but deviate more than MLPs at increased and decreased volumes as well as BCC structures (Supplementary Figure 2).
Energy minima predicted by the potential of Winey et al.~\cite{winey2009thermodynamic} are shifted to lower volumes, indicating underestimated lattice parameters.

Similar to the occurrence of different crystal structures, dislocations and grain boundaries emerge during quenching and evolve during mechanical deformation.
To ensure these are depicted accurately, we investigate the predicted stacking fault energy (SFE) $\gamma$ along the normalized displacement $d$ (Figure \ref{fig:crystal}b).
The SFE is evaluated by incrementally displacing the top half of an FCC block along the ${111}$ slip plane in the $\langle112\rangle$ direction (Supplementary Note 3).
SFEs calculated using GNNP-Al and ANI-Al consistently remain close to DFT calculations, even though dislocated structures are not explicitly present in either ANI-Al or LE datasets.
The non-refined GNNP-Al and UMA models deviate the most from the reference (Figure \ref{fig:crystal}b).
They specifically underestimate the stable fault energy $\gamma_{sf}$ at $d=1/3$, where the FCC packing locally resembles an HCP structure~\cite{anderson2017theory}.
This agrees with the results from the energy evaluations (Figure \ref{fig:crystal}a).
Beyond $\gamma_{sf}$, the two maxima of the fault curve influence defect formation during solidification.
The first peak corresponds to the unstable stacking fault energy $\gamma_{us}$ and represents the barrier for nucleation of a leading Shockley partial dislocation~\cite{christian1995deformation, borovikov2016effects}.
Reduced $\gamma_{us}$, as observed for UMA and GNNP-Al (non-refined), facilitates dislocation nucleation, resulting in higher defect densities and earlier yielding (Figure \ref{fig:deformation_ovito} and \ref{fig:deformation_plot}, Supplementary Table 5 and 6).
The classical potentials by Lee et al.~\cite{lee2003semiempirical} and Mendelev et al.~\cite{mendelev2008analysis}, on the other hand, overestimate this barrier (Supplementary Figure 2), leading to fewer defects and thus a higher yield stress.
The second peak corresponds to the unstable twin fault energy $\gamma_{ut}$ and governs the emission of the trailing partial~\cite{christian1995deformation}.
If this barrier is underestimated, as for the potentials by Lee et al.~\cite{lee2003semiempirical} and Winey et al.~\cite{winey2009thermodynamic} (Supplementary Figure 2), twinning occurs more frequently, promoting twin thickening and resulting in either narrow stacking faults or multilayer HCP regions associated with deformation twins (Figure \ref{fig:deformation_ovito}, Supplementary Figure 11)~\cite{tadmor2004first, zhang2019stacking}.

We further investigated solid properties at increasing temperatures by computing the elasticity tensor $C$ and the lattice constant.
Simulation setups and evaluation procedures for all solid-state properties are described in Supplementary Note 3.
Due to the symmetries of the FCC crystal, only three of the 81 components of the elasticity tensor $C$ are independent, namely $C_{11}$, $C_{12}$, and $C_{44}$.
$C_{11}$ captures the stiffness in the tensile direction and therefore strongly influences the Young’s modulus during deformation, while $C_{12}$ describes lateral coupling effects, which are typically less significant.
Finally, $C_{44}$ characterizes shear behavior and plays a key role in grain slip and material yielding.

\begin{figure}[!htbp]
    \centering
    \includegraphics[width=.8\linewidth]{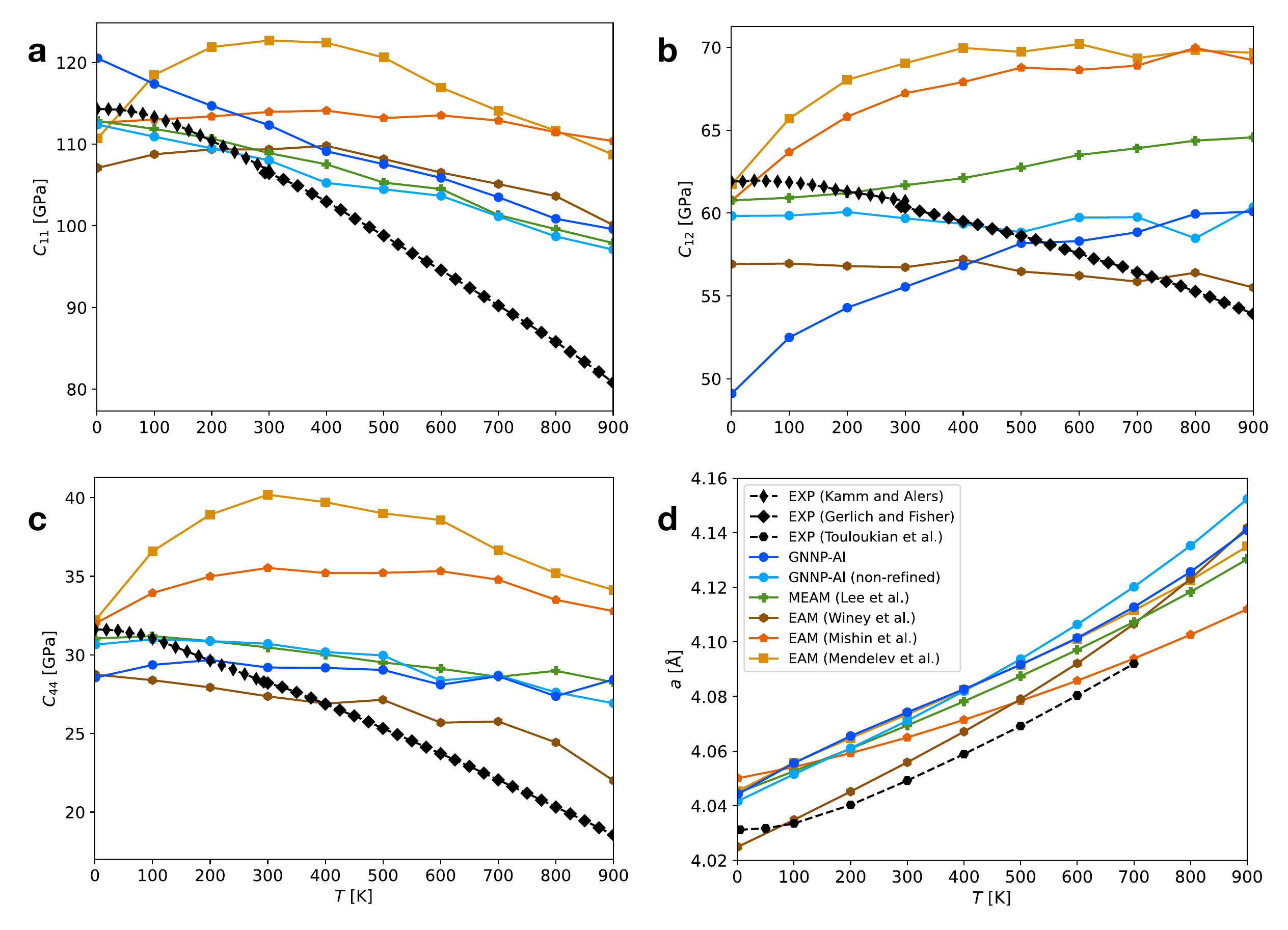}
    \caption{Elasticity tensor components $C_{11}$ (a), $C_{12}$ (b), and $C_{44}$ (c) and lattice parameter $a$ (d) of an FCC crystal as a function of temperature $T$ across the solid range. Experimental values are taken from Kamm and Alers~\cite{kamm1964low} and Gerlich and Fischer~\cite{gerlich1969high} for the low and high temperature elasticity components, respectively. For the lattice parameter, experimental values correspond to work by Touloukian et al.~\cite{touloukian1975thermal}.}
    \label{fig:solid_state_properties}
\end{figure}

Overall, the GNNP-Al model predicts a mechanical response similar to that of the MEAM potential (Figure \ref{fig:solid_state_properties}a-c).
Both show better agreement with experiments than the classical EAM models of Mendelev et al.~\cite{mendelev2008analysis} and Mishin et al.~\cite{mishin1999interatomic}, which consistently overestimate all elastic constants.
This behavior can be attributed to insufficient anharmonic softening in the EAM functional, which leads to inaccurate higher-order derivatives and an incorrect temperature dependence.
As a result of their elevated elastic constants, simulations using these models overestimate the material’s stiffness, underpredict grain slip, and delay the onset of yield (Figure \ref{fig:deformation_plot}b, Supplementary Figure 7, Supplementary Table 5 and 6).
The opposite trend is observed for the potential of Winey~\cite{winey2009thermodynamic}, which was fitted to extrapolated room-temperature measurements rather than to DFT calculations using energy-minimized structures. Consequently, this potential also somewhat underestimates the values at absolute zero relative to the other classical potentials.

Interestingly, at lower temperatures, GNNP-Al performs slightly worse than the non-refined GNNP-Al model, even though it was specifically trained on near-ideal crystal data.
This effect is most pronounced for $C_{12}$ and likely arises because the refinement steps alter the local curvature in low-energy states.
Since elastic constants are second-order derivatives of the energy, refinement based on only a few energy and force samples may improve the energy prediction in this region while distorting the curvature, thereby degrading the elastic response.
Nevertheless, at the temperatures considered in this work (above 300 K), both models agree well with each other and with the experimental references.
Moreover, perfect agreement with experiment is not expected, as DFT itself shows deviations from experimental results, and similar discrepancies were found for other MLPs trained on DFT data~\cite{rocken2024accurate}.

Similarly, we observe small discrepancies between model predictions and experimental results for the lattice parameters (Figure \ref{fig:solid_state_properties}d). Specifically, predictions are consistently about 1\% higher than the experimental thermal expansion measurements~\cite{touloukian1975thermal} across the entire temperature range. This discrepancy arises from the PBE functional, which tends to overestimate lattice parameters~\cite{spagnoli2010density} and has been used to generate the ANI-Al dataset and to fit many classical potentials. The only exception is the EAM model developed by Winey et al.~\cite{winey2009thermodynamic} which was fitted to extrapolated experimental data. Despite this, all predictions remain within the 1\% error margin of PBE and can therefore be considered accurate across all models. The solid-state properties predicted by the other GNNP-Al models are presented in Supplementary Figure 4.

\subsection{Liquid Properties}
\label{sec:liquid}

\begin{figure}[!htbp]
    \centering
    \includegraphics[width=.8\linewidth]{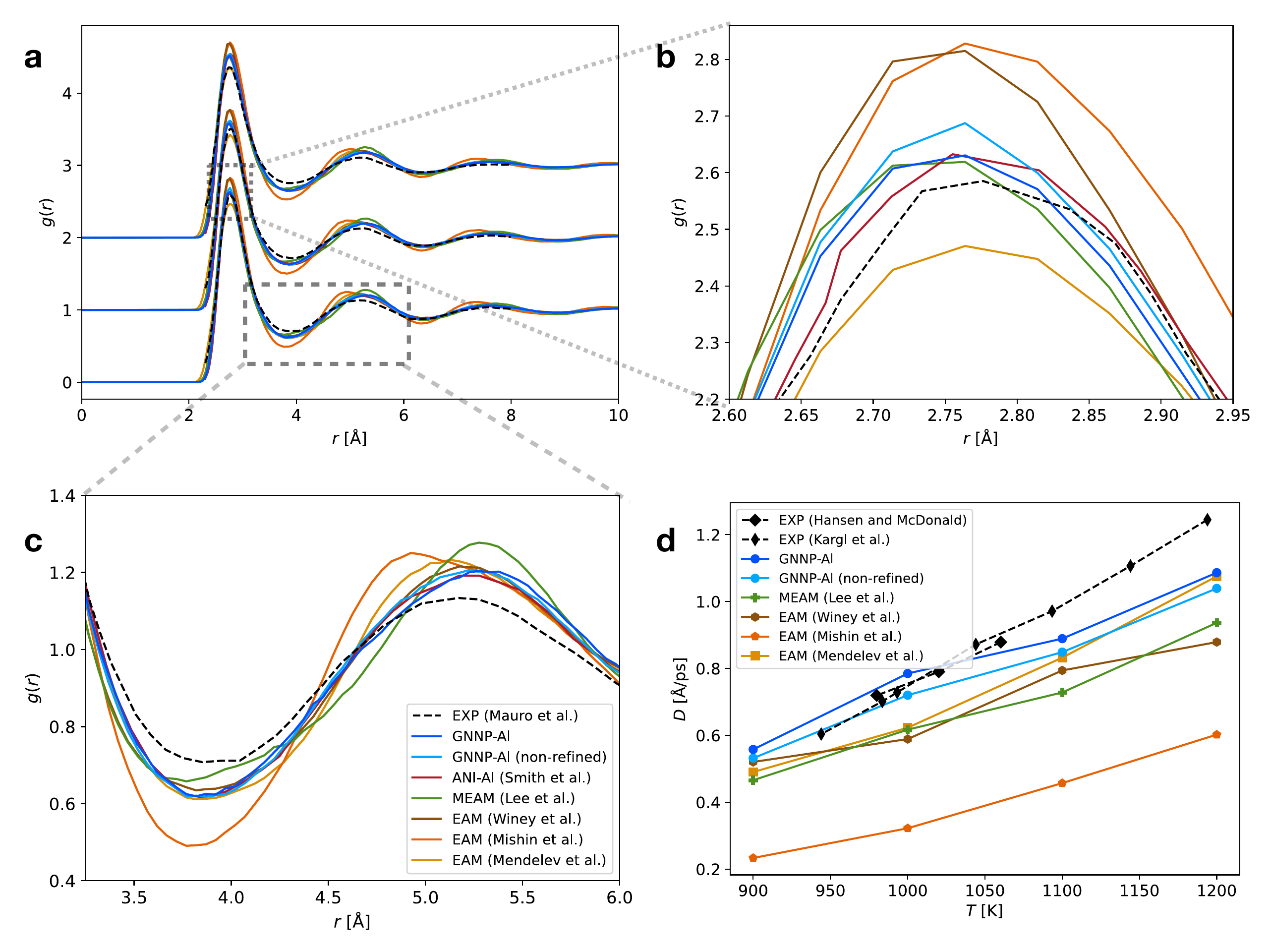}
    \caption{(a) Radial distribution function (RDF) $g(r)$ at three different temperatures $T$, 1123 K, 1183 K, and 1273 K. RDFs at increasing temperatures are shifted up by a value of one to distinguish between temperatures. Close-ups of the RDF at 1123 K are provided for the first peak (b), as well as the region of the first valley and second peak (c). Experimental RDFs are taken from Mauro et al.~\cite{mauro2011high}. Predictions for ANI-Al were taken from Smith et al.~\cite{smith2021automated}.
    (d) Diffusion coefficient $D$ of liquid aluminum plotted against temperature. Experimental values are from Kargl et al.~\cite{kargl2013impact} as well as Hansen and McDonald~\cite{hansen2013theory}.}
    \label{fig:liquid_properties}
\end{figure}

For the molten metal, we assess the radial distribution function (RDF) and the diffusion coefficient over a broad temperature range above the melting point.
Both properties are evaluated from 500 ps NVT simulations.
The RDF is computed using the Open Visualization Toolkit (OVITO)~\cite{stukowski2009visualization}, and we derive the diffusion coefficient from the temporal gradient of the mean square displacement.
The complete simulation setup and workflow are described in Supplementary Note 4.
In the liquid state, our trained GNNP-Al and the ANI-Al model strongly agree with experiments and outperform all tested classical potentials (Figure \ref{fig:liquid_properties}).
All potentials predict qualitatively similar RDFs, but differences arise due to the various approaches to capturing temperature dependencies.
This effect is especially pronounced at the peaks, where the models that are not explicitly fitted to liquid properties, such as those by Mishin et al.~\cite{mishin1999interatomic} and Winey et al.~\cite{winey2009thermodynamic}, exaggerate the peaks, i.e., modeling the fluid closer to a solid, than the other models by Mendelev et al.~\cite{mendelev2008analysis} and Lee et al.~\cite{lee2003semiempirical}, which included properties such as the melting temperature during fitting (Figure \ref{fig:liquid_properties}b and c).
RDFs predicted by the MEAM potential of Lee et al.~\cite{lee2003semiempirical} are similar to those predicted by GNNP-Al and ANI-Al at short distances, but deteriorate at larger ones (Figure \ref{fig:liquid_properties}c), underscoring the limited expressivity of the model compared to MLPs.

The superiority of GNNP-Al over classical potentials for the liquid state is further highlighted by the results for the diffusion coefficient, which is underestimated by all classical potentials to varying extents (Figure \ref{fig:liquid_properties}d, Supplementary Figure 5d).
For solidification, this poses a significant challenge, as diffusion affects the kinetic prefactor of the nucleation rate~\cite{sosso2016crystal}.
Low diffusion can lead to a vanishing nucleation rate and, therefore, no grain formation.
We observe this effect in this work for the case of the potential by Mishin et al.~\cite{mishin1999interatomic}, which caused solidification into an amorphous glass structure (Figure \ref{fig:solidification_plot}).
We can further observe that a lower diffusion coefficient correlates with a delayed onset of grain formation in the examples of the models by Winey et al.~\cite{winey2009thermodynamic} and Lee et al.~\cite{lee2003semiempirical} (Figure \ref{fig:solidification_plot}).

\subsection{Solid-Liquid Coexistence}
\label{sec:coexistence}

The melting temperature best quantifies the solid-liquid phase transition.
To determine it, we performed three independent coexistence simulations (see Supplementary Note 5 and Supplementary Figure 6 for details).
Although GNNP-Al predicts the structural properties of the liquid with high accuracy, it underestimates the experimental melting temperature (Table \ref{tab:melt_temp}). 
\begin{table}[!htbp]
    \centering
    \begin{tabular}{l r}
        \toprule
        Model & $T_m [\si{K}]$ \\
        \midrule
        Experiment & 933 \\
        \addlinespace[0.5ex]
        GNNP-Al & 877.2 ($\pm$0.4) \\
        GNNP-Al (non-refined) & 818.4 ($\pm$1.2) \\
        \addlinespace[0.5ex]
        MEAM (Lee et al.~\cite{lee2003semiempirical}) & 914.2 ($\pm$1.1) \\
        EAM (Winey et al.~\cite{winey2009thermodynamic}) & 817.3 ($\pm$1.2) \\
        EAM (Mishin et al.~\cite{mishin1999interatomic}) & 1018.5 ($\pm$2.1) \\
        EAM (Mendelev et al.~\cite{mendelev2008analysis}) & 906.2 ($\pm$0.7) \\
        \bottomrule
    \end{tabular}
    \caption{Melting temperature $T_m$ as predicted by different potentials via the coexistence method. The standard deviation from the ensemble of three runs is reported in brackets.}
    \label{tab:melt_temp}
\end{table}
Nevertheless, the deviation remains within a reasonable range and does not exceed 10\%.
Interestingly, the GNNP-Al (non-refined) model predicts a substantially lower melting temperature, even though the refinement incorporated only ideal-crystal information, highlighting the sensitivity of melting-temperature predictions to small changes in MLPs.
Previous studies have likewise reported that MLPs can underpredict melting temperatures even by several hundred kelvin~\cite{fuchs2025refining}.
Classical potentials fitted to thermal properties, such as those of Mendelev et al.~\cite{mendelev2008analysis} and Lee et al.~\cite{lee2003semiempirical}, predict melting temperatures closer to experiment, whereas those that were not, predict larger deviations than GNNP-Al.

Differences in melting temperature across models are smaller than the corresponding differences in diffusion and therefore have a less pronounced effect on the solidification process.
Even so, the melting temperature strongly influences the nucleation rate~\cite{sosso2016crystal}.
A higher predicted melting temperature leads to greater supercooling at a given temperature.
Whether this increases or decreases the nucleation rate depends on competing effects, which control the critical supercooling temperature~\cite{sosso2016crystal}.
In the quenching simulations reported here, the supercooling exceeded this critical value.
Therefore, a higher melting temperature resulted in a lower nucleation rate (Supplementary Figure 8, Supplementary Table 3).

\subsection{Solidification}
\label{sec:solidification}

\begin{figure}[!htbp]
    \centering
    \includegraphics[width=.8\linewidth]{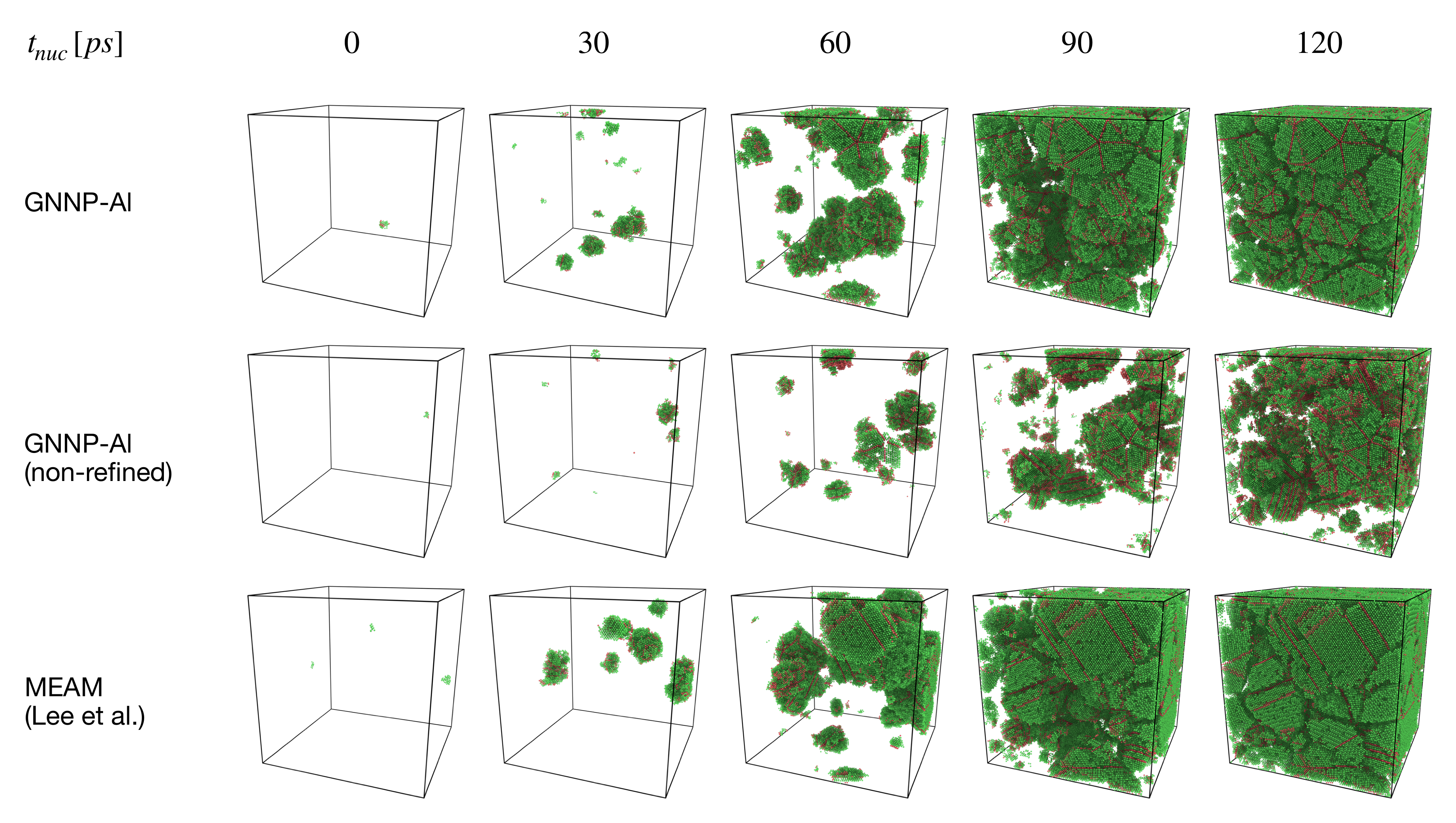}
    \caption{Time evolution of solidification using three different simulation potentials. The leftmost frame marks the first grain formation ($t_{nuc}$). The time difference between the images is 30 ps. Atoms colored in green are assembled in a face-centered cubic (FCC) structure, while red atoms correspond to hexagonal close-packing (HCP). Amorphous structures are deleted for grain visibility.}
    \label{fig:solidification_ovito}
\end{figure}

After extensively analyzing the fundamental properties, we proceeded to solidification simulations involving one million atoms. Specifically, a cubic box containing $1,000,188$ atoms was quenched at a rate of $1 \times 10^{12}$ K/s to induce solidification. 
Simulation details are provided in Supplementary Note 6, and the computational performance of each model is discussed in Supplementary Note 7 and summarized in Supplementary Table 4.
Visual inspection of the resulting structures reveals the formation of twin boundaries and five-fold twins in the GNNP-Al and MEAM simulations (Figure \ref{fig:solidification_ovito}). 
Such features commonly arise during rapid solidification~\cite{hofmeister1998forty, wang2001characterization, zhu2005formation, an2011formation, qin2015recoverable, zhang2017formation} and have also been reported in comparable MEAM simulations by Mahata and Asle Zaeem~\cite{mahata2019evolution}. 
The emergence of similar defect structures in both the GNNP-Al and MEAM simulations suggests that the two models capture comparable solidification dynamics.
In contrast, the EAM potentials deviate from this behavior and primarily produce twin boundaries without five-fold twins (Figure \ref{fig:solidification_ovito}, Supplementary Figure 11). 
This difference arises from EAM's potential dependency on only pairwise distances and therefore lack of explicit angular contributions, unlike MEAM and GNNP-Al models. 
Since angular distortions play a critical role in stabilizing five-fold twins, inaccuracies in their description can render five-fold junctions unstable.

\begin{figure}[!htbp]
    \centering
    \includegraphics[width=.8\linewidth]{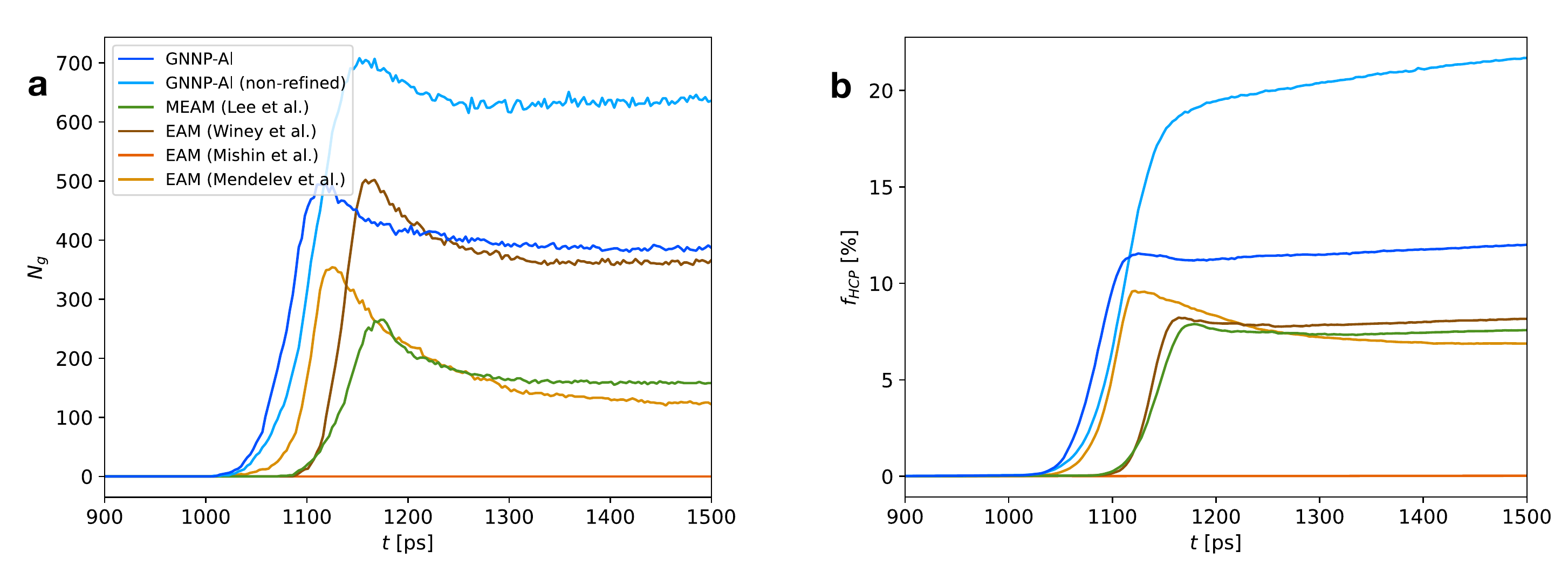}
    \caption{Number of grains $N_g$ (a) and fraction of atoms ordered in hexagonal close-packed (HCP) structure $f_{HCP}$ (b) over time $t$ during quenching simulations. The results are shown only from 900 ps after the start of quenching, because no nucleation was observed before that time in any simulation.}
    \label{fig:solidification_plot}
\end{figure}

We quantify the solidification process by tracking both the number of grains and the fraction of atoms in HCP stacking during the simulation.
First, the predicted stacking fault energies strongly influence the resulting grain structure.
The non-refined GNNP-Al model predicts a significantly larger HCP fraction, consistent with its reduced stacking fault energy and low unstable twin energy (Figure \ref{fig:crystal} and \ref{fig:solidification_plot}).
As a result, more atoms adopt HCP stacking and stacking faults are increasingly identified as grain boundaries, leading to a higher number of grains.
We observe this correlation between HCP fraction and grain count for all investigated potentials (Figure \ref{fig:solidification_plot}).
A similar behavior would therefore be expected for the UMA potential, which exhibits comparable energetic deviations as the GNNP-Al (non-refined) model.

Second, the diffusion coefficient strongly affects the nucleation process.
In the simulation using the EAM potential by Mishin et al.~\cite{mishin1999interatomic}, no grains formed because the severely underestimated diffusion suppressed nucleation.
The limited atomic mobility in the liquid prevented the formation of crystalline nuclei, resulting in solidification into an amorphous glass.
Consistent with this interpretation, the potentials by Lee et al.~\cite{lee2003semiempirical} and Winey et al.~\cite{winey2009thermodynamic}, which also predict low diffusion coefficients, show a delayed onset of nucleation (Figure \ref{fig:liquid_properties}d and \ref{fig:solidification_plot}). 
A similar dependence is observed for the melting point. For example, the GNNP-Al models that predict higher melting temperatures display a slower onset of solidification (Supplementary Figure 8, Supplementary Table 3).

Finally, stochastic effects primarily influence the later stages of solidification.
Because nucleation and grain growth are inherently stochastic processes, we performed three simulations with identical parameters but different initial velocity seeds (Supplementary Figure 9).
Despite this stochasticity, the onset of nucleation occurs nearly simultaneously across all runs, as it is governed by the nucleation rate.
In contrast, the subsequent grain merging differs across simulations (Supplementary Figure 9), reflecting the random spatial distribution and growth rates of individual nuclei.
These observations indicate that while stochasticity affects the later stages of microstructure evolution, the early stages of solidification are largely determined by the physical properties predicted by the potential.

\subsection{Deformation}
\label{sec:deformation}

To evaluate the impact of the potential on the predicted material behavior, we further perform tensile tests on the solidified structures. We compute mechanical properties, including Young's modulus, yield strength, and ultimate tensile strength.
Details of the tensile tests and the evaluation of the mechanical properties are provided in Supplementary Note 8 and visualized in Supplementary Figure 10.

\begin{figure}[!htbp]
    \centering
    \includegraphics[width=.6\linewidth]{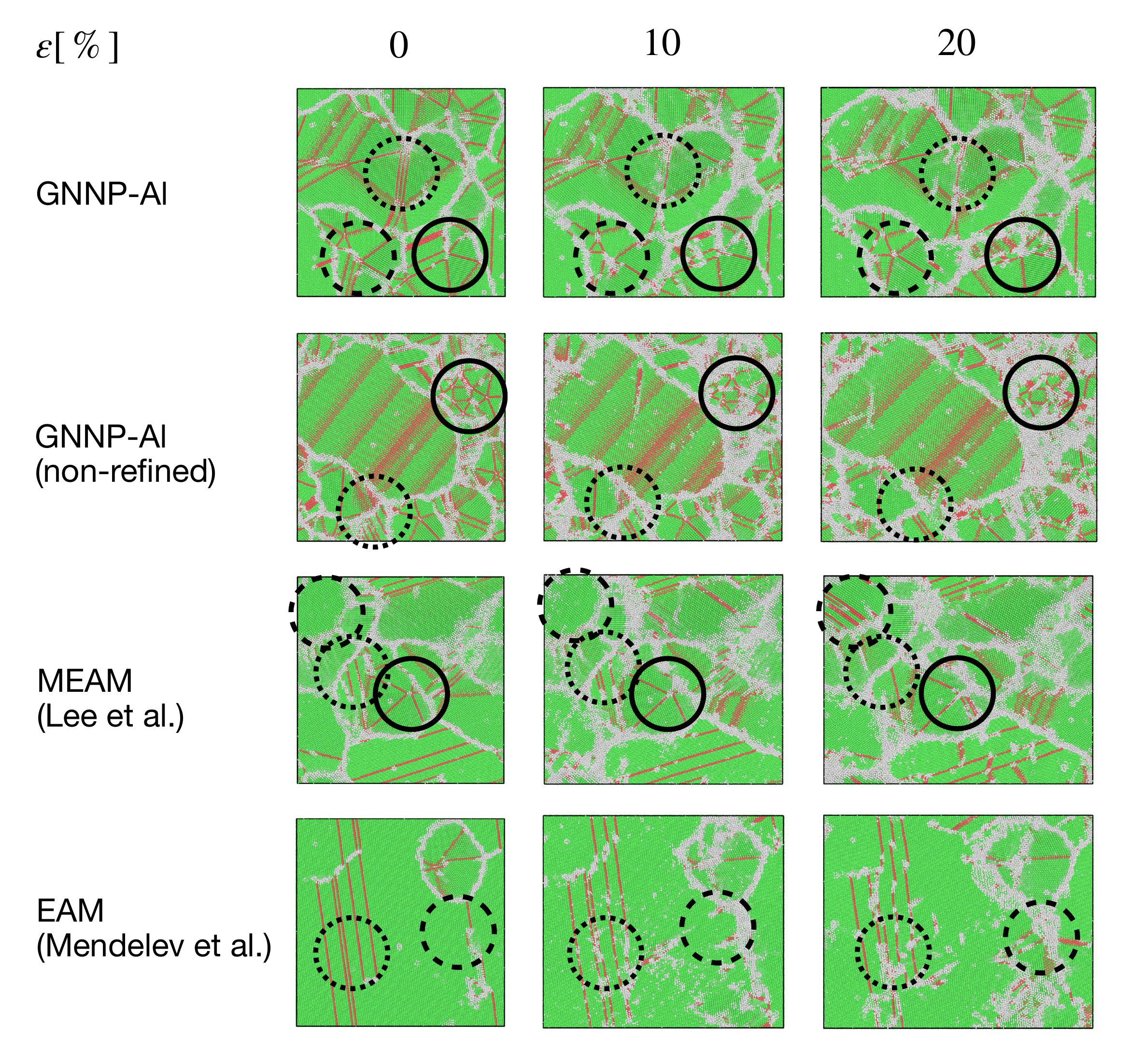}
    \caption{Grain structures during the tensile test in $x$-direction at three different stages. At $\varepsilon=0$, the structure corresponds to the undeformed one after quenching. Two more states are added at 10\% and 20\% deformation. All figures are from a top view, i.e., showing the $xy$-plane. Circles point to areas where typical changes in the grain structure occur. Solid circles show the dispersion of five-fold twins. Dashed and dotted ones indicate twinning and detwinning under deformation, respectively. Atoms colored in green are packed according to the FCC structure, those in red belong to HCP crystals, and gray atoms are part of an amorphous structure. }
    \label{fig:deformation_ovito}
\end{figure}

The evolution of the grain structure during deformation is first analyzed visually.
Previously reported phenomena, such as twinning and detwinning, are observed for all models (Figure \ref{fig:deformation_ovito}, Supplementary Figure 11)~\cite{mahata2019evolution, li2009reversible, kibey2007predicting, yamakov2002deformation}.
Models that underestimate the unstable twin energy $\gamma_{ut}$, such as those by Lee et al.~\cite{lee2003semiempirical} and Winey et al.~\cite{winey2009thermodynamic} (Supplementary Figure 2), exhibit enhanced slip activity and consequently show more pronounced twin thickening and detwinning compared to the other models (Figure \ref{fig:deformation_ovito}, Supplementary Figure 11).
For clarity, only a single surface is shown for each case.
However, additional surfaces and cross-sectional views of the simulation boxes reveal consistent behavior.

\begin{figure}[!htbp]
    \centering
    \includegraphics[width=.8\linewidth]{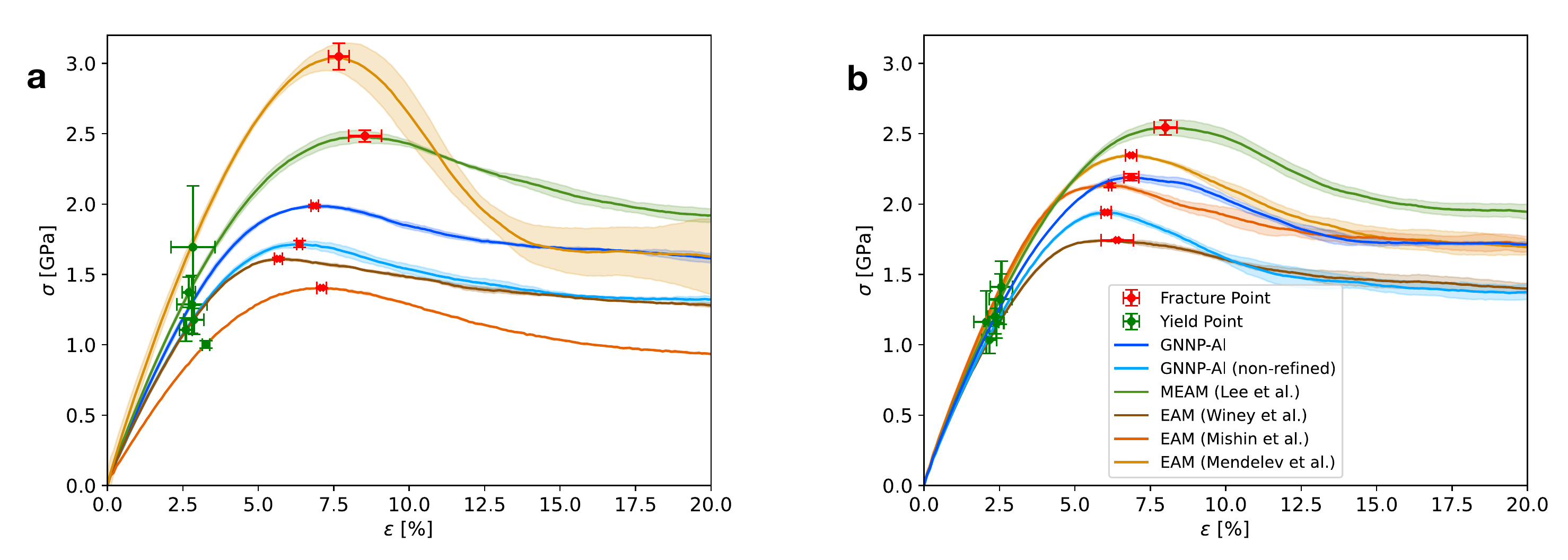}
    \caption{Stress $\sigma$ vs. strain $\epsilon$ plots showing mechanical responses under tensile deformation simulated with different models. The structure after quenching using the same model (a) is stretched up to 20\% at a strain rate of $\dot{\varepsilon} = 10^{10}$ 1/s. The same is repeated with all models using the same input structure generated by the GNNP-Al model (b). Stress responses are averaged over all directions. Yield and fracture points are marked with green and red crosses, respectively. Error bars and bands represent one standard deviation of the three runs in the different directions.}
    \label{fig:deformation_plot}
\end{figure}

To complement the visual analysis, we quantify the deformation behavior using stress–strain curves.
All models produce curves with a similar overall shape, characterized by an initial elastic regime, subsequent yielding, and eventual fracture (Figure \ref{fig:deformation_plot}).
While this indicates qualitatively consistent deformation mechanisms across all potentials, the extent of the individual regimes and the resulting mechanical properties vary substantially between models.
The applied potential influences the tensile response in two ways.
First, it governs the atomic interactions during deformation.
Second, it determines the microstructure of the initial configuration generated during the quenching simulation.
To isolate the potential's effect during deformation alone, we repeated the tensile tests using identical initial structures but different interaction models.
Specifically, we selected the structure obtained from a quenching simulation with the GNNP-Al potential as the reference configuration.
The subsequent deformation simulations were performed under the same conditions as the original tensile tests (Supplementary Note 7).

The increased strength of the structure solidified with the potential by Mendelev et al.~\cite{mendelev2008analysis} originates from the presence of a few large FCC grains, while the increased variance arises from parallel dislocations within the structure (Figure \ref{fig:deformation_plot}, Supplementary Figure 8 and 10).
In contrast, the amorphous structure produced by the potential of Mishin et al.~\cite{mishin1999interatomic} reduces both the strength and the anisotropy of the mechanical response.
The predictions obtained for the structure solidified with GNNP-Al deviate less strongly from the other models, indicating that the resulting microstructure, and therefore the solidification process, can substantially influence the mechanical properties.
Consistently, the remaining potentials show similar deformation characteristics for both structures on which they were tested (Figure \ref{fig:deformation_plot}).
The model by Winey et al.~\cite{winey2009thermodynamic} predicts a low Young's modulus as well as early yielding and fracture, consistent with its low predicted elastic constants and unstable twinning energy (Figure \ref{fig:crystal}, \ref{fig:solid_state_properties}, and \ref{fig:deformation_plot}; Supplementary Tables 5 and 6).
Similarly, the potentials by Mendelev et al.~\cite{mendelev2008analysis} and Mishin et al.~\cite{mishin1999interatomic}, which both overestimate elastic constants, predict a large Young's modulus.
Finally, models predicting a high unstable stacking fault energy, such as those by Mendelev et al.~\cite{mendelev2008analysis} and Lee et al.~\cite{lee2003semiempirical}, lead to an increased yield stress.
These observations emphasize that inaccuracies in ideal crystal properties propagate into both solidification and deformation simulations, highlighting the importance of accurately reproducing these properties when developing new interatomic potentials.

\begin{figure}[!htbp]
    \centering
    \includegraphics[width=.8\linewidth]{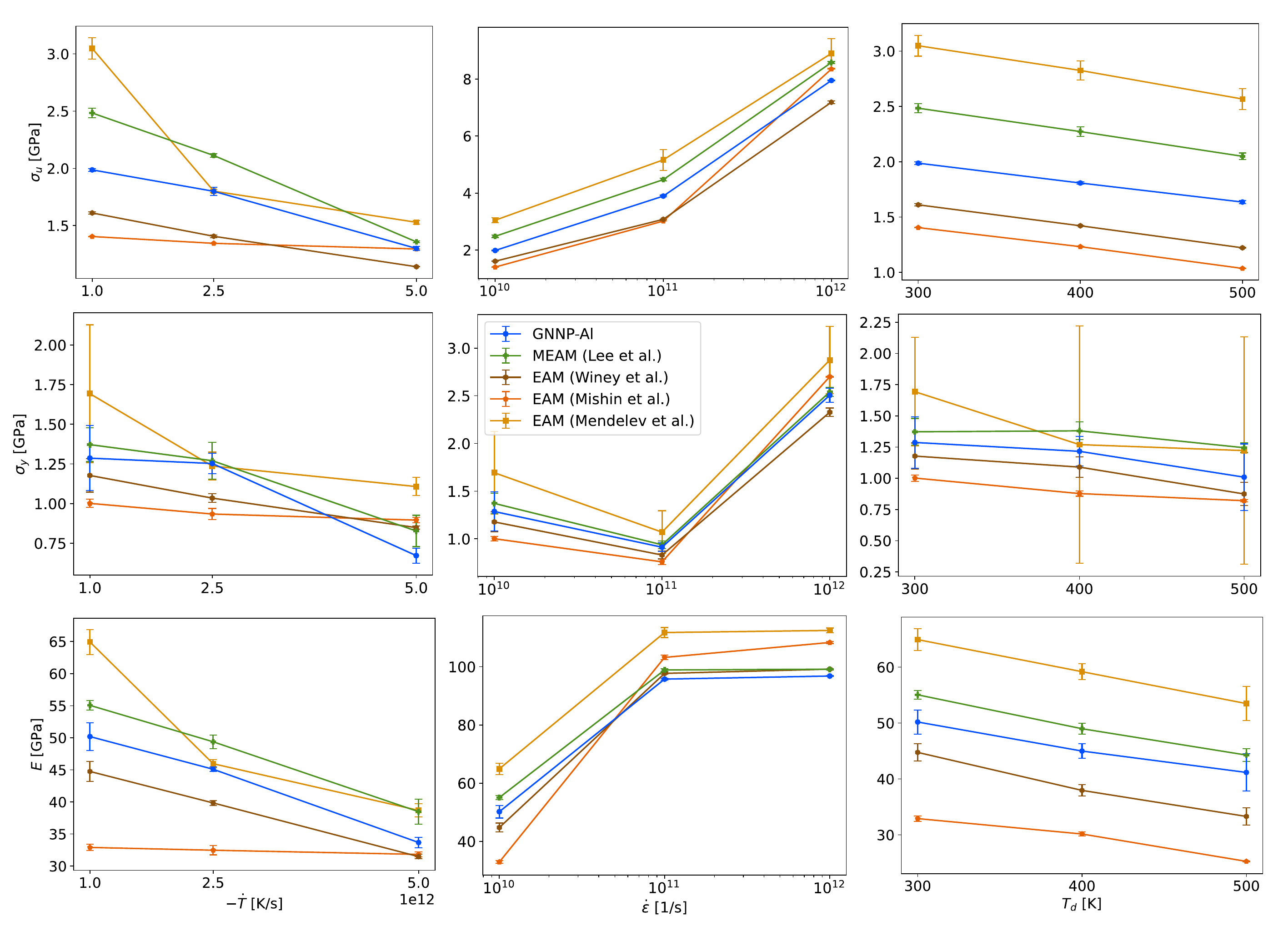}
    \caption{Dependencies of three key mechanical properties (bottom to top), Young's modulus $E$, yield strength $\sigma_y$, and ultimate tensile strength $\sigma_u$ over three properties (left to right), quench rate $-\dot{T}$, strain rate $\dot{\epsilon}$, and deformation temperature $T_d$. Error bars represent one standard deviation of the three runs in the different directions.}
    \label{fig:dependencies}
\end{figure}

Finally, we investigated the influence of several simulation parameters on the mechanical properties.
Specifically, we compare three values for the quench rate, strain rate, and deformation temperature.
In each tensile test, only one parameter is varied, while all other conditions remain identical to those used in the previous simulations.
All models exhibit qualitatively similar responses to the changing simulation conditions.
In particular, all recorded strength metrics decrease with increasing quench rate and deformation temperature (Figure \ref{fig:dependencies}, Supplementary Figure 13).
A higher quench rate leads to a higher defect density and residual stresses, which negatively affect mechanical properties.
Similarly, higher deformation temperatures soften the material through thermal expansion, thereby facilitating slip and dislocation motion.
In contrast, increasing the strain rate results in a stronger mechanical response because deformation mechanisms have less time to activate, leading to strain-rate hardening.
These trends are consistent with previous simulations using a MEAM potential reported by Mahata and Asle Zaeem~\cite{mahata2019effects}.

\section{Conclusion}

In summary, we have developed the GNNP-Al model, an MLP for aluminum based on a state-of-the-art GNN architecture.
We demonstrated its improved fidelity compared to classical force fields, lower memory requirements than ANI-Al MLP~\cite{smith2021automated}, and an order-of-magnitude lower computational cost than the foundational UMA model~\cite {wood2025family}.
Taken together, these improvements highlight GNNP-Al's relevance for studying solidification and deformation phenomena that require accurate models across a wide temperature regime and simulations spanning large spatiotemporal scales. 

Through a detailed analysis of the performance of several classical and MLP models, we highlighted key material properties that strongly affect solidification processes.
In particular, we find that while GNNP-Al predicts diffusion in good agreement with experiments, the classical potentials underestimate it by 15~\% to as much as 50~\%, which significantly impacts nucleation rate.
Similarly, GNNP-Al (non-refined) and the foundational UMA model failed to clearly distinguish competing crystal structures and underestimated the unstable stacking fault energy by more than 50~\%, leading to large deviations in phase fractions and grain counts.
In contrast, all models predict the melting temperature within 10~\% of the experimental reference.
Thus, correlations between melting-point error and the solidification process are not as prominent.
Evaluation of mechanical properties revealed that overestimated unstable twin-fault energies increased yield strength, underscoring the importance of matching stacking-fault energy across the entire displacement spectrum. On the other hand, incorrect modeling of ideal-crystal elasticity components propagates to the Young's modulus of a more complex solidified nanopolycrystal.

Using this array of observations, we could devise a novel methodology for training MLPs via sequential refinement, where the MLP is initially trained on an active-learning-generated dataset and later refined on low-energy states.
Active learning is an established method for generating MLP datasets and can exploit various strategies to generate and select the most informative configurations for training~\cite{smith2021automated, smith2018less, podryabinkin2017active, thaler2024active}.
Most commonly, and in the case of the ANI-Al dataset, high-temperature MD sampling is combined with ensemble uncertainty quantification~\cite{smith2021automated, zhang2020dp}.
Active learning generally increases the applicability of MLPs across a wide temperature range, thereby improving liquid-state accuracy and MD stability.
Nevertheless, the underrepresentation of low-energy structures in the dataset and the emphasis the RMSE loss places on high-energy states can lead to a deficient modeling of solid-state properties, as we showed for the GNNP-Al (non-refined) model.
Interestingly, similar deficiencies are observed in UMA, suggesting that sequential refinement could be an effective approach to fine-tuning foundational models~\cite{wong2026bias}.
Presently, we only exploit ideal crystal structures with varying lattices during refinement.
However, the addition of stacking fault structures would better align the model with the underlying ab initio predictions.

To further improve the MLP's accuracy beyond that of DFT, top-down learning, in which the MLP is trained to match experimental observations, can be employed.
Such techniques are at the core of classical potential development and were also used to parametrize the EAM and MEAM models explored in this work.
While classical models can be adjusted via trial and error as they contain only a few adjustable parameters, MLPs require systematic approaches.
The liquid-state radial distribution function and, in principle, diffusion could be matched using automatic differentiation~\cite{schoenholz2020jax, greener2025reversible}.
Nevertheless, differentiating through the MD simulation is computationally and memory-intensive.
Thus, methods such as DiffTRe~\cite{thaler2021learning, han2025refining}, which exploit reweighting, are needed to match properties, requiring more extensive sampling.
For example, substantial discrepancies with experiments for the lattice parameters and the elasticity tensor that we and others report can be minimized~\cite{rocken2024accurate}.
Furthermore, recent work also demonstrated that MLPs can be fine-tuned to reproduce the experimental phase diagram~\cite{fuchs2025refining, swinburne2025agnostic}. 

While we employed a GNNP here to study the solidification and deformation of pure aluminum, the same methodology can be directly applied to multicomponent systems.
Since, contrary to Behler-Parinello type MLPs, GNNPs scale favorably with the number of atom types, we expect that solidification of multicomponent systems could be accurately simulated at a similar computational cost to that presented here, thus paving the way for a deeper understanding of solidification.

\section*{CRediT} 
\textbf{Ian Störmer:}
Conceptualization, Methodology, Software, Formal analysis, Investigation, Visualization, Writing - Original Draft.
\textbf{Julija Zavadlav:}
Conceptualization, Validation, Writing - Review \& Editing, Resources, Funding acquisition, Supervision, Project Administration.

\section*{Acknowlegements}

Funded by the Deutsche Forschungsgemeinschaft (DFG, German Research Foundation) – Projektnummer: 534045056.

The authors gratefully acknowledge the Gauss Centre for Supercomputing e.V. (www.gauss-centre.eu) for funding this project by providing computing time on the GCS Supercomputer JUWELS~\cite{JUWELS} at Jülich Supercomputing Centre (JSC).

\section*{Data and Code Availability}

The code and data supporting this study will be made publicly available on GitHub upon acceptance of this manuscript.

\bibliographystyle{ieeetr}
\bibliography{references}

\end{document}